\documentclass[preprint,floatfix,showpacs,superscriptaddress]{revtex4}
\usepackage{graphicx}
\usepackage{verbatim}
\usepackage{psfrag}
\usepackage{natbib}
\begin{document}
\def\baru{\bar{u}}
\def\phiav{\phi}
 
\title{Turbulence collapses at a threshold particle loading in a 
dilute particle-gas suspension.}

\author{V. Kumaran}
\affiliation{Department of Chemical Engineering, Indian Institute
of Science, Bangalore 560 012, India.}
\author{P. Muramalla}
\affiliation{Department of Chemical Engineering, Indian Institute
of Technology Bombay, Mumbai 400 076, India.}
\author{A. Tyagi}
\affiliation{Department of Chemical Engineering, Indian Institute
of Science, Bangalore 560 012, India.}
\author{P. S. Goswami}
\affiliation{Department of Chemical Engineering, Indian Institute
of Technology Bombay, Mumbai 400 076, India.}

\begin{abstract}
Two mechanisms are considered responsible for the turbulence modification 
due to suspended particles in a turbulent gas-particle suspension. Turbulence
augmentation is due to the enhancement of fluctuations by wakes behind particles,
whereas turbulence attenuation is considered to result from the increased
dissipation due to the particle drag. In order to examine the turbulence
attenuation mechanism, Direct Numerical Simulations (DNS) 
of a particle-gas suspension are carried out at a
Reynolds number of about 3333 based on the average gas velocity $\bar{u}$,
channel width $h$, and the gas kinematic viscosity. The 
particle Reynolds number based on the particle diameter $d_p$, gas kinematic
viscosity and the flow velocity $\bar{u}$ is about 42 and the 
Stokes number is in the range $7-450$. The particle volume
fraction is in the range $0-2 \times 10^{-3}$, and the particle mass loading
is in the range $0-9$. As the volume fraction is increased, a discontinuous
decrease in the turbulent velocity fluctuations is observed at a critical 
volume fraction. 
There is a reduction, by one order of
magnitude, in the mean square fluctuating velocities in all directions and 
in the Reynolds stress. Though there is a modest increase in the 
energy dissipation due to particle drag, this increase is smaller than 
the decrease in the turbulent energy production; moreover, there is a 
decrease in the total energy dissipation rate when there is turbulence 
collapse.
Thus, turbulence attenuation appears to be due to a 
disruption of the turbulence production mechanism, and not due to 
the increased dissipation due to the particles.
There is a discontinuous collapse in the turbulence intensities 
at a critical particle loading, instead of
the continuous decrease as the particle loading is increased.
\end{abstract}


\maketitle   

A central issue in the dynamics of particle-gas suspensions is the effect
of the particles on the gas phase turbulence. This issue is of great
physical significance in geophysical phenomena such as sandstorms and
snow avalanches, as well as in industrial applications such as fluidised
beds and pneumatic transport. 
An important question, which has been studied extensively, is whether
the fluid turbulence is increased or decreased due to the presence
of the particles.

When the suspended
particles are in the size range $10-100 \mu$m, the particle Reynolds
number (ratio of fluid inertia and viscosity based on the 
particle diameter) is small, and so the drag force on the particles
can be adequately described by the Stokes law or a modified drag
law that incorporates inertial corrections. The particle Stokes
number (ratio of fluid inertia and particle viscosity) is typically
large. Due to this, the particles cross the fluid 
streamlines due to inertia, and there is a force exerted on the fluid 
due to the instantaneous
difference in the particle and fluid velocities. 

An early review by Gore and Crowe\cite{Gore_1989} 
reported that turbulence modification is determined by the 
ratio $(d_p/L)$, where $d_p$ is the particle diameter and $L$
is the integral length scale of turbulence. Turbulence was found
to be suppressed for $(d_p/L) < 0.1$, and augmented for $(d_p/L)
> 0.1$. 
Direct Numerical Simulations of particle-laden turbulent flows 
by Squires \& Eaton\cite{Squires_1991}, found that particles
are preferentially concentrated in high strain regions, and expelled
from high vorticity regions. Elghobashi \& Turesdell\cite{Elghobashi_1993} 
found that particles increase the turbulent energy at high wave numbers, 
and suggested a transfer of energy from the large to small scales
due to the particles in the absence of gravity, and a reverse cascade
from small to large scales in the presence of gravity. 
Li and McLaughlin\cite{Mclaughlin_2001} reported 
that particles increase the turbulence intensity at low loading, whereas
they suppress the turbulence intensity at high loading. Based on a
review of experimental and numerical results, Tanaka \& Eaton
\cite{Tanaka_2008}
identified a particle momentum number; they found that particles
with small momentum number increase turbulence, whereas those with
moderate momentum number decrease turbulence. Recent 
simulations of Vreman\cite{Vreman_2015} and
Capecelatro et al\cite{capecelatro_2018} also report
that the turbulence intensity progressively
decreases with particle loading, as the increased dissipation
due to the particles compensates for the decrease in the fluid
turbulent energy production. 


Here, we critically examine
the turbulence attenuation mechanism, specifically whether the decrease
in turbulence intensity is continuous as the particle loading is increased,
and whether the excess dissipation due to the particle phase does result
in turbulence attenuation. The Reynolds number based on the average flow
velocity, channel width and the gas kinematic viscosity is set to a fixed
value of $3333$. The particle Reynolds number $\mbox{Re}=(\rho d_p
\Delta u / \mu)$ is $42$, where $\rho$ is the fluid density, $d_p$ 
is the particle diameter, $\Delta u$ is the maximum difference in the 
mean particle and fluid velocities and $\mu$ is the gas viscosity. 
The particle Stokes number is the ratio of the viscous relaxation time
and the fluid time scale, the latter here considered as $\tau_f = (h/\bar{u})$.
For the inertia-corrected
drag law, equation \ref{eq:drag}, the viscous relaxation time is
$\tau_v = (\rho_p d_p^2 / 18 \mu (1 + 0.15 \mbox{Re}^{0.667}))$.
The ratio of the two time scales, $\mbox{St} = (\tau_v / \tau_f)$ is
varied in the range 
$1.91-153.16$. The particle volume fraction is varied in the 
range $0-2 \times 10^{-3}$, resulting in a variation in the range $0-9$
for the particle mass loading. The Kolmogorov scale, $(\nu^3/
\epsilon)^{1/4}$, is about $1.6 \times 10^{-2}$ times the channel 
width for the unladen flow, where $\epsilon$ is the rate of dissipation
of energy due to the turbulent fluctuations, and $\nu = (\mu/\rho)$
is the kinematic viscosity. The particle diameter is comparable to the
Kolmogorov scale, so we use the point force approximation 
for the particles. This also implies that the particles can not directly
disrupt the coherent structures in the flow.

\begin{figure}
 \includegraphics[width=.25\textwidth]{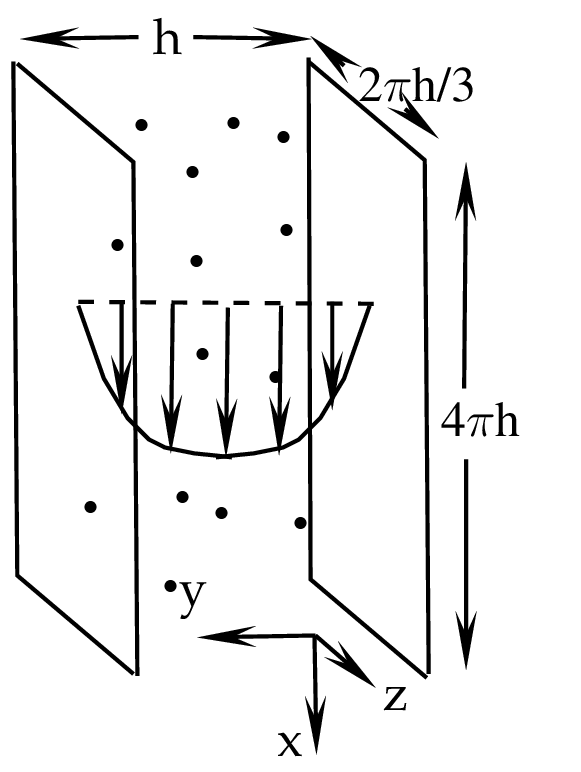}
 \caption{\label{fig:config} Configuration and co-ordinate system used for the simulations.}
\end{figure}

The configuration and co-ordinate system used for the simulations is
shown in figure \ref{fig:config}. The dimensions of the channel are
$4 \pi h \times h \times (2 \pi h / 3)$ in the flow (x), wall-normal (y)
and the span-wise (z) directions, where $h$ is the channel width. Zero
velocity boundary conditions are applied at the walls $y = \pm h/2$, 
while periodic boundary conditions are applied at in the flow and
the span-wise directions. Spectral DNS is used, where Fourier transforms
are used in the stream-wise and span-wise directions which are periodic,
and Chebyshev transforms are used in the cross-stream direction with
zero velocity boundary conditions. In order to resolve the smallest
scales at $\mbox{Re}=3333$, 128 Fourier modes are used in the flow
direction, $64$ Fourier modes in the span-wise direction and
$65$ Chebyshev modes in the cross-stream ($y$) direction.

The particles are considered to be rigid spheres, 
and the ratio of the particle diameter
and the channel width is $1.84 \times 10^{-2}$. The particle terminal
velocity is $4 \times 10^{-3}$ smaller than the fluid average velocity,$(St = 38.3)$ and
so the gravitational force on the particles is negligible compared to 
the drag force due to the fluid. 
The particle Reynolds number is too large for the Stokes law to 
be valid, and so an inertia-corrected drag law \cite{dritselis_2016},
\begin{equation}
{\bf F}_I^D = - 3 \pi \mu d_p ({\bf u} - {\bf v}_I) (1 + 
0.15 \mbox{Re}_I^{0.667}),
\label{eq:drag}
\end{equation}
is used. Here, ${\bf F}_I^D$ is the drag force on particle $I$,
${\bf v}_I$ is the particle velocity, ${\bf u}$ is the fluid
velocity at the particle location, $d_p$ is the particle diameter, 
and $\mbox{Re}_I$ is the particle Reynolds
number based on the particle diameter, difference in gas and
particle velocities $|{\bf u} - {\bf v}_I|$ and the gas kinematic 
viscosity. Equation \ref{eq:drag} is accurate to within about 2\% for 
$\mbox{Re} = 42$ for an isolated particle in a fluid.

The hard sphere molecular dynamics simulation procedure is used for the 
particles, where the particle positions are updated based on the 
particle velocity, and the particle velocity is evolved using 
Newton's laws. The force on the particle is the sum of the 
gravitational force (which is negligible in the present case), 
the drag force (equation \ref{eq:drag}) and the force due to 
inter-particle and particle-wall collisions. The inter-particle
collisions are modeled using the elastic hard-sphere model,
where the relative velocity between a pair of particles along
the line joining centers is reversed in a collision, while the 
relative velocity perpendicular to the line joining centers is
unchanged. The particle-wall collisions are considered specular,
where the particle velocity perpendicular to the wall is 
reversed in a collision, while the particle velocity parallel to
the wall is unchanged.
The force exerted by the particle on the fluid, which is the negative
of the drag force, is treated as a delta function force located at 
the particle center. This approximation is justified because the 
particle diameter is comparable to the Kolmogorov length 
scale. This force is included in the Chebyshev-Fourier
transformed fluid momentum equations. 

The initial condition for the simulation is a steady unladen
turbulent flow, where the particles are added at random locations
with the same velocity as the fluid velocity. The simulation is
run for $3000$ integral times to reach steady state, where the 
integral time is the ratio of the channel width and the mean 
fluid velocity. The fluid and particle statistics are then calculated
over a period of $1000$ integral times. The mean fluid velocity,
$\bar{u}_x$, is a function of the cross-stream
co-ordinate $y$. The velocity fluctuations in the flow,
cross-stream and span-wise directions are $u_x', u_y'$ and
$u_z'$. The overbars are used to denote time averages, for example
the mean square velocity in the flow direction is $\overline{u_x'^2}$.
The cross-section averaged flow velocity $\bar{u}$ and the 
channel width $h$ in the cross-stream direction are used
for non-dimensionalisation.

\begin{figure}
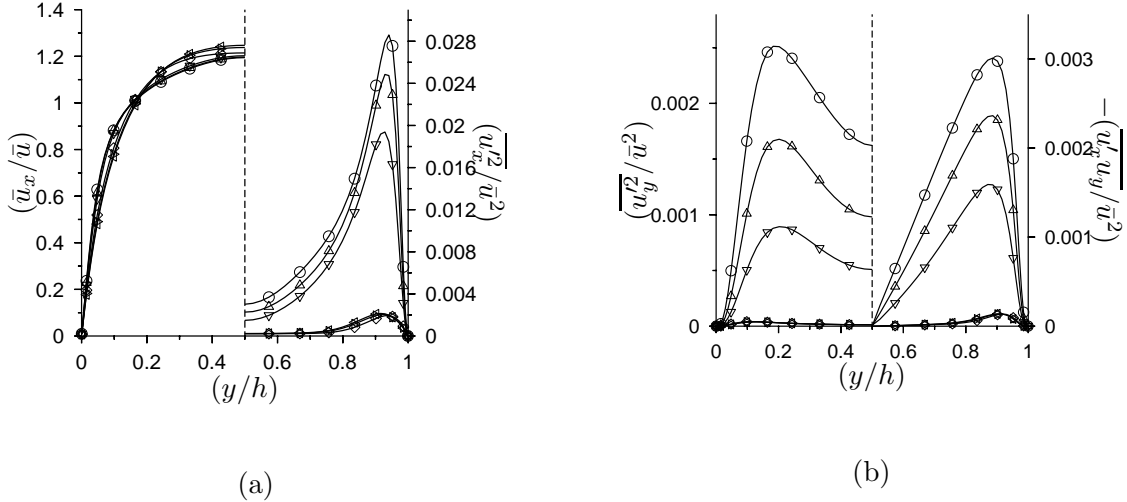

\parbox{.49\textwidth}{ 
  \psfrag{x}[][][1][0]{$(y/h)$}
  \psfrag{y}[][][1][0]{$(\bar{u}_x/\baru)$}
  \psfrag{yr}[][][1][0]{$(\overline{u_x'^2}/\baru^2)$}
   \includegraphics[width=.4\textwidth]{prlfig1.ps}

\begin{center} (a) \end{center}
}
\parbox{.49\textwidth}{
  \psfrag{x}[][][1][0]{$(y/h)$}
  \psfrag{y}[][][1][0]{$(\overline{u_y'^2}/\baru^2)$}
  \psfrag{yr}[][][1][0]{$- (\overline{u_x' u_y}/\baru^2)$}
   \includegraphics[width=.4\textwidth]{prlfig2.ps}
   
\begin{center} (b) \end{center}
}
\caption{\label{fig:fig1} In sub-figure (a), the fluid mean velocity 
(left half) and the stream-wise mean square fluctuating velocity 
(right half), and in sub-figure (b), the cross-stream mean square
velocity (left half) and the correlation $\overline{u_x' u_y'}$
(right half), all suitably
scaled by the powers of the average flow velocity $\baru$,
as a function of the scaled co-ordinate
$(y/h)$, for average particle volume fraction $0$ ($\circ$), $5 \times 10^{-4}$ 
($\triangle$), $9 \times 10^{-4}$ ($\nabla$), $10^{-3}$ 
($\triangleleft$), $1.1 \times 10^{-3}$ ($\triangleright$), and
$1.4 \times 10^{-3}$ ($\diamond$). The particle Stokes number is
$94.5$.
}
\end{figure}

The effect of particle loading on the mean and the root mean square
of the fluctuating velocities are shown in figures \ref{fig:fig1}
for particle Stokes number $38.3$.
The mean and root mean square velocities are symmetric about
the center-line of the channel, and so each quantity is plotted
in one half of the figure. The left half of
figure \ref{fig:fig1} (a) shows the variation in
mean velocity profiles with particle volume fraction. 
The mean velocity is close to the turbulent velocity
profile for the unladen turbulent flow when the volume fraction
is increased from $0$ to $9 \times 10^{-4}$. There is a distinct
change in the velocity profile when the volume fraction is increased
from $9 \times 10^{-4}$ to $10^{-3}$. The velocity profile has
a lower curvature near the walls and smaller gradient at the walls.
When the volume fraction is further increased to $1.4 \times 10^{-3}$,
there is little change in the velocity profile. 

The change in the mean velocity profile is accompanied by a drastic
reduction in the mean square velocities in all three directions.
The stream-wise mean square velocity in the right half of
figure \ref{fig:fig1} (a)
exhibits the characteristic near-wall maximum for a turbulent flow
when the volume fraction is $9 \times 10^{-4}$ or less. When the 
volume fraction is increased from $9 \times 10^{-4}$ to 
$10^{-3}$, there is a dramatic collapse in the mean square
of the stream-wise fluctuations by one order of magnitude. Upon further
increase in the volume fraction, there is very little change in the 
mean square of the fluctuating velocities.

A dramatic decrease is also observed
in the mean square velocities in the wall-normal direction, as shown 
in the left half of figure \ref{fig:fig1} (b). More importantly, 
there is also a virtual collapse in the Reynolds stress as shown in the 
right half of figure \ref{fig:fig1} (d). This implies that the suspension 
stress is primarily due to viscosity and momentum transport by the 
particles for volume fraction $10^{-3}$ and above.

The mean square of the particle velocity fluctuations have also been
examined. The particle concentration and mean velocity are uniform
across the channel, and there is virtually no change in the particle
mean velocity as the volume fraction of the particles is increased.
 There is a small but discernible increase of about 10-20\%
in the mean square velocities of the particles, but there is no 
dramatic change as that observed in the fluid turbulence. 

\begin{figure}
  \psfrag{y}[][][1][0]{Critical volume fraction}
  \psfrag{x}[][][1][0]{St}
\includegraphics[width=.5\textwidth]{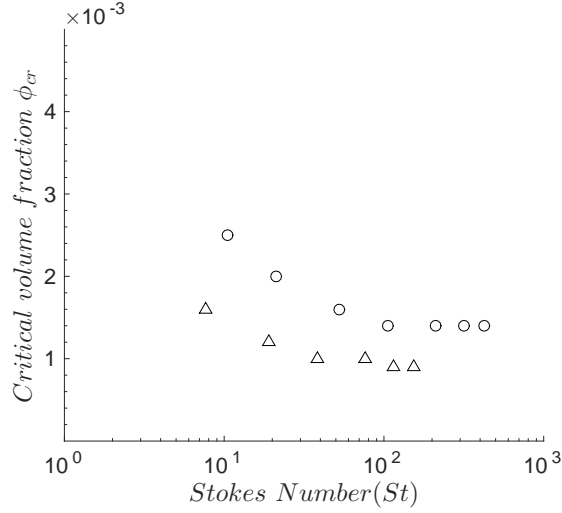}
\caption{\label{fig:criticalvf} The critical volume fraction for the 
turbulence collapse
transition as a function of the particle Stokes number (St) for the 
inertia-corrected drag law equation \ref{eq:drag} ($\triangle$) and the 
Stokes drag law $(\circ)$.
}
\end{figure}

The discontinuous transition in the turbulence intensities is observed 
for other values of the particle Stokes number, and also
if the Stokes drag law is used instead of the inertia-corrected
drag law equation \ref{eq:drag}. 
This indicates that the discontinuous transition is a 
robust process independent of the particle Stokes number and the 
details of the drag law used. The critical volume fraction, is
shown as a function of the particle Stokes number  in figure 
\ref{fig:criticalvf}. For Stokes drag law with inertial correction, critical volume fraction is independent
of Stokes number when $St \ge 40$, and it appears to increase as the Stokes number decreases for $St \le 40$.
Similarly for Stokes drag law, the critical volume fraction increase as the Stokes number decreases when $St \le 100$ 
and then it is constant for $St \ge 100$ 

In order to examine the mechanism of turbulence attenuation, we have
calculated separately the total rate of dissipation of fluid kinetic
energy, the rate of dissipation due to the particle drag ${\cal D}_p$,
and the rate of turbulent production of energy ${\cal P}$. 
The scaled rate of production of kinetic energy across 
the channel, which is also the scaled dissipation rate
due to the turbulent velocity fluctuations at steady
state, is,
\begin{eqnarray}
{\cal P} & = & \mbox{} - \frac{h}{\bar{u}^3} \langle \overline{u_x'
u_y'} \frac{d \bar{u}_x}{d y} \rangle_s,
\end{eqnarray}
where $\langle \rangle_s$ is the spatial average across the channel.
The scaled rate of dissipation of energy due to the mean shear is,
\begin{eqnarray}
{\cal D}_f & = & \frac{\mu}{(\rho \bar{u}^3/h)} \langle
\left( \frac{d \bar{u}_x}{d y} \right)^2 \rangle_s.
\end{eqnarray}
The scaled rate of dissipation of energy due to the 
drag force exerted by the particles is 
\begin{eqnarray}
{\cal D}_p & = & \frac{h}{\rho \bar{u}^3 V}
\sum_I \overline{{\bf u \cdot F}_I},
\end{eqnarray}
where ${\bf F}_I$ is the force exerted by particle $I$ on the fluid, and
$V$ is the total volume. The total rate of dissipation of 
energy per unit mass of gas ${\cal D} = {-\cal P} + {\cal D}_p + {\cal D}_f$.

\begin{figure}
  \psfrag{x}[][][1][0]{Volume fraction}
  \psfrag{y}[][][1][0]{${\cal P}, {\cal D}, {\cal D}_p, {\cal D}_f$}
   \includegraphics[width=.5\textwidth]{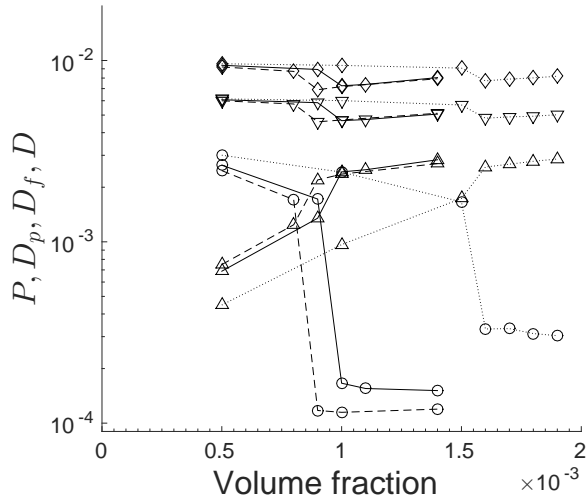}

\caption{\label{fig:keterms} The scaled total rate of dissipation of energy
per unit mass ${\cal D}$ ($\diamond$), the rate of transfer of energy per 
unit mass from the mean flow to the fluctuations ${\cal P}$ ($\circ$),
the rate of
dissipation of energy due to the drag force exerted on the particles
${\cal D}_p$ ($\triangle$), and the rate of dissipation of energy due to
the mean shear in the fluid ${\cal D}_f$ ($\nabla$) as a function of volume 
fraction for particle Stokes number $7.66$ (dotted lines),
$38.3$ (solid lines), $114.9$ (dashed lines).
}
\end{figure}

The rates of total energy dissipation, turbulent production and 
dissipation due to the particles are shown as function of the 
volume fraction for three different values of the particle Stokes
number in figure \ref{fig:keterms}. It is observed that there
is a significant decrease even in the total rate of dissipation of
energy at the critical volume fraction. There is a dramatic collapse
in the rate of turbulent production. There is an increase in the 
rate of dissipation due to the particles, but this increase is 
only about one half of the decrease in the turbulent production.
Thus, turbulence collapse is not accompanied by a compensatory
increase in the energy dissipation due to the particle drag,
but instead there is a decrease in the total energy dissipation rate.
This indicates that the turbulence attenuation is due to a 
disruption of the turbulence production mechanism, rather than
an increase in dissipation due to the particles.

The present study has been limited to a relatively low Reynolds number
of about 3333. The other parameters have been varied over relatively
wide ranges --- two different drag laws have been used, the Stokes
drag law and a modification of the Stokes law with inertial
correction, equation \ref{eq:drag}, six different values of the particle
Stokes number has been studied for each drag law, about 5-6 particle
loadings have been considered for each Stokes number to detect 
turbulence collapse, and each simulation has been repeated at least
3 times resulting in a comprehensive study involving more than 200
simulations.  
This study has uncovered a heretofore unknown phenomenon, which is 
the discontinuous decrease in the turbulence intensity at a critical
volume loading in a particle-gas suspension. This is in contrast to 
the conventional wisdom that there is a gradual decrease in the
turbulence intensities as the particle loading is
increased. The mechanism for turbulence
modification, the disruption of the turbulence production of
energy in the gas phase, is different from the mechanism of increased
particle dissipation which was previously considered responsible for
turbulence attenuation. In the present case,
there is no comparable increase in the dissipation due to the 
particle fluctuations, and therefore the turbulence attenuation
is not due to the increased dissipation by the particles.

The authors thank the SERB, Department of Science and Technology,
Government of India, for financial support. VK would like to thank
the JRD Tata Trust for supporting this research.

\bibliographystyle{unsrt}
\bibliography{prl_ref}

\end{document}